\documentclass[conference]{IEEEtran}
\IEEEoverridecommandlockouts
\usepackage{cite}
\usepackage{amsmath,amssymb,amsfonts}
\usepackage{algorithmic}
\usepackage{graphicx}
\usepackage{tabularx}
\usepackage{algorithm}
\usepackage{algorithmic}
\usepackage{textcomp}
\usepackage{listings}
\usepackage{xcolor}

\colorlet{punct}{red!60!black}
\definecolor{background}{HTML}{EEEEEE}
\definecolor{delim}{RGB}{20,105,176}
\colorlet{numb}{magenta!60!black}

\lstdefinelanguage{json}{
    basicstyle=\normalfont\ttfamily,
    showstringspaces=false,
    backgroundcolor=\color{background},
    basicstyle=\small,
    literate=
     *{0}{{{\color{numb}0}}}{1}
      {1}{{{\color{numb}1}}}{1}
      {2}{{{\color{numb}2}}}{1}
      {3}{{{\color{numb}3}}}{1}
      {4}{{{\color{numb}4}}}{1}
      {5}{{{\color{numb}5}}}{1}
      {6}{{{\color{numb}6}}}{1}
      {7}{{{\color{numb}7}}}{1}
      {8}{{{\color{numb}8}}}{1}
      {9}{{{\color{numb}9}}}{1}
      {:}{{{\color{punct}{:}}}}{1}
      {,}{{{\color{punct}{,}}}}{1}
      {\{}{{{\color{delim}{\{}}}}{1}
      {\}}{{{\color{delim}{\}}}}}{1}
      {[}{{{\color{delim}{[}}}}{1}
      {]}{{{\color{delim}{]}}}}{1},
}

\def\BibTeX{{\rm B\kern-.05em{\sc i\kern-.025em b}\kern-.08em
    T\kern-.1667em\lower.7ex\hbox{E}\kern-.125emX}}
\begin{document}

\title{Open source platform Digital Personal Assistant
\\
}

\author{\IEEEauthorblockN{Azat Khusnutdinov}
\IEEEauthorblockA{
\textit{Innopolis University}\\
Innopolis, Russia \\
a.khusnutdinov@innopolis.ru}
\and
\IEEEauthorblockN{Denis Usachev}
\IEEEauthorblockA{
\textit{Innopolis University}\\
Innopolis, Russia \\
d.usachev@innopolis.ru}
\and
\IEEEauthorblockN{Manuel Mazzara}
\IEEEauthorblockA{
\textit{Innopolis University}\\
Innopolis, Russia \\
m.mazzara@innopolis.ru}
\and
\IEEEauthorblockN{Adil Khan}
\IEEEauthorblockA{
\textit{Innopolis University}\\
Innopolis, Russia \\
a.khan@innopolis.ru}
\and
\IEEEauthorblockN{Ivan Panchenko}
\IEEEauthorblockA{
\textit{Innopolis University}\\
Innopolis, Russia \\
i.panchenko@innopolis.ru}
}
\maketitle

\begin{abstract}

Nowadays Digital Personal Assistants (DPA) become more and more popular. DPAs help to increase quality of life especially for elderly or disabled people. In this paper we develop an open source DPA and smart home system as a 3-rd party extension to show the functionality of the assistant. The system is designed to use the DPA as a learning platform for engineers to provide them with the opportunity to create and test their own hypothesis. The DPA is able to recognize users' commands in natural language and transform it to the set of machine  commands that can be used to control different 3rd-party application. We use smart home system as an example of such 3rd-party. We demonstrate that the system is able to control home appliances, like lights, or to display information about the current state of the home, like temperature, through a dialogue between a user and the Digital Personal Assistant.

\end{abstract}

\begin{IEEEkeywords}
Personal Assistant, Smart Home, Conversation Support, Command recognition
\end{IEEEkeywords}

\section{Introduction}
Nowadays we live in times when society has been changing the way of interaction with machines. Thanks to the achievements in the field of Natural Language Processing (NLP) and Machine Learning (ML), computers can better and better recognize natural languages. One of the applications of this technology was found in a Digital Personal Assistant. 
\par Personal Assistant is a software that helps users to achieve goals by performing tasks and services on behalf of an individual. To be efficient, a DPA usually combines user input with other implicit information that it can gather, such as information from online sources, location, and history of previous user behavior. There are a lot of different Personal Assistants that were created for the last ten years, the most popular ones are Siri, Cortana and Alexa. 
\par The user can interact with a Personal Assistant via different interfaces, but most common ones are textual commands through chat and voice. Textual and voice commands require from developers to match user input to executable commands. It requires solving a huge list of NLP problems such as lemmatization, part of speech tagging, chunking, named entity recognition, intent extraction, support dialog and so on. After the extraction of the intent from user's query, a DPA can delegate the task to other 3rd party services that will execute commands \cite{Action_on_Google, SiriKit, Mycroft_skills, Jasper_module}.
\par The creation of a complex DPA that is able to handle a huge variety of tasks is very hard and time-consuming and only a few big companies like Microsoft or Google are able to create such types of assistants \cite{sarikaya2016overview}. These assistant are commercial products thus they do not provide any access to their source code, which means developers are limited in being able to modify them and test hypothesis. 
\par One of the trading applications of Personal Assistant is the control of the smart home environments. Smart homes and smart building provide access to sophisticated monitoring and control over buildings functions. Controlling different properties of the building such as temperature, light, humidity as well as some more complicated factors like presence, identity recognition and even emotional states of the people \cite{Encyclopedia_of_E_Health} becomes easier every year, mainly because of decreasing price of hardware components and increasing power of microprocessors \cite{khanda2017microservice}. These hardware components allowing us to control the environment and collect the data are usually referred as Internet of Things (IoT).
\par In this paper, we introduce our Open Source DPA that is capable to understand user's intent, extract parameters from utterances, transform them to machine commands and route the commands to Smart Home automation module that handles the execution of these commands. The system can be easily extended by other modules developed by 3rd party developers. 
\par In the following chapters, we are going to provide the general background of existing work, describe the architecture of the system and its requirements. Then we will present results that we obtained and talk about the current and future steps.

\section{Related Work}

\subsection{Personal Assistant}

Digital Personal Assistants are very popular nowadays, thanks to Google, Amazon and Apple more and more people start using DPA’s in their everyday life \cite{Tractica}. A modern Personal Assistant does not only performs actions on behalf of a user but also tries to predict the user’s actions and provide proactive assistance. Many early assistants were focused on specific tasks, but nowadays more and more systems can be extended to deal with a variety of different tasks. We can observe it in the examples of Google's Assistant \cite{Action_on_Google} and Apple's Siri \cite{SiriKit}. Such architecture makes a DPA more and more intelligent and user friendly.
\par But everything has a price. DPAs involve more and more computationally expensive methods of data processing in order to achieve human capabilities in understanding a text, speech and recognition of objects in images. A team from the University of Michigan has made DPA “Lucida” with the aim to evaluate requirements to future data centers which play the role of back system for processing DPA tasks  \cite{hauswald2015sirius}. A team of engineers demonstrated that the voice query requires 164 times greater time than traditional query in a web searcher. Such time consumption imprints on the scalability of the system. As the information volume grows, the system spends more time on applying different NLP filters in order to choose the best answer of a question. Developers provide specific measurements of time consumption for a Question-Answering module which spends on average 85\% of the cycles on three components: stemming, regular expression pattern matching and CRF. In the paper of J. Hauswald et al. demonstrate possible solutions related to the problems of resource consumption, but still Lucida can not be run on modern smartphones.

\par Modern DPAs are moving towards an understanding and supporting dialog with the user. R. Sarikaya et. al. present an architecture of Cortana developed with the aim to realize a practical system capable to answer questions from any domain dialog. They introduce 10 problems of optimization for DPA, for example the complexity of a dialog, the naturalness of the user language, computational cost and so on \cite{sarikaya2016overview}.

\subsection{Comparison of Personal Assistants}
Nowadays high level architecture of all DPAs resemble each other. We can highlight modules such as Speech-To-Text, Text Processing, Information extraction, Intent handlers and Text-To-Speech. Variability of actions which a DPA can perform depends on the plugged intent handlers \cite{Action_on_Google, SiriKit, Mycroft_skills, Jasper_module}. For this reason, comparing assistants by plugged modules does not reflect the real quality of them. The main advantages of a DPA are hidden in natural language processor modules. Companies that own proprietary assistants do not provide detail technical information about their implementation of the modules. In its turn, open source DPAs such as MyCroft, Open Assistant, Jasper and ADRIAN use open source libraries and some of the libraries can be integrated in several assistants, e.g. library Pocketsphinx is involved in MyCroft, Open Assistant and Jasper \cite{Open_Assistant_Pocketsphinx, Mycroft_Pocketsphinx, Jasper_Pocketsphinx}. For this reason, we should compare no entire DPAs to another one, but analyze each module separately. But this comparison is beyond the scope of our article.

\subsection{Natural language processing}
Each language is characterised by great diversity. People can express their opinions with several different proposals in writing, but with the same meaning. For example, if you want to change the temperature at your home you can say "Set up temperature at 25 degree" or "I feel hot, limit heater till 20". For this reason, having a system capable to understand rephrases is extremely important. One of the ways to solve this task is measuring text similarity. Scientists develop new and new methods of measure. One of them is Word Mover’s Distance (WMD) \cite{kusner2015word}. The approach is based on word embedding. The WMD distance measures the dissimilarity between two text documents as the minimum amount of distance that the embedded words of one document need to “travel” to reach the embedded words of the other. M. Kusner et al demonstrate the power of the approach on two examples in different documents: “Obama speaks to the media in Illinois” and “The President greets the press in Chicago”. While these sentences have no words in common, they convey nearly the same information. This fact cannot be represented by the bag-of-words model, but can be caught by WMD.

\subsection{IoT and Smart Homes}
For the last few years, there has been a huge growth of Internet Of Things \cite{IOT_Platforms_Market_Report}. There are a lot of definitions for IoT \cite{b9,santos2016intelligent}, but it is usually used to refer to a set of objects that are connected to the internet and can possibly communicate with each other via different protocols. These can be various sensors, actuators, embedded electronics and etc. At the moment there are around 22 billion \cite{b11} IoT devices all over the world. Internet Of Things paradigm enables us to create a huge amount of different communication scenarios, where all IoT devices are supposed to communicate and cooperate \cite{tan2010future,mekikis2016information}. Different scenarios could be used in different fields of our life, such as in Home automation, Health, Transportation and Logistics and more \cite{b9}.
\par A smart home consists of a set of IoT devices communicating with each other to achieve the goals of home automation and monitoring. The market has many different IoT devices, but unfortunately, not all of them are compatible with each other and with a Digital Personal Assistant. Managing the communication between the devices is a big challenge when building home automation  \cite{khanda2017microservice}. Another problem is that most of IoT devices are not capable of managing computationally intensive tasks, thus centralized control is required in order to build complex Automation system.
\par  Smart homes are automated environments that provide users with access to sophisticated monitoring and control over the house environment. The idea of building smart home appeared in 1970s  \cite{b21}, but the costs of automation were high and required significant time and effort. Usually smart homes operate with light, temperature, air conditioning and security, but they can be extended to control any part of the house like multimedia and even kitchen appliances and etc. Since prices and sizes of micro-controllers significantly have been decreasing for the last 10 year we can automate almost anything that people can think of. Home automation can be divided into three layers \cite{b22} by the tasks they are intended to solve:
\begin{itemize}
    \item Sensing - provides a system with information from all the different sensors deployed in the home. These data provide an understanding of the current state of the house.
    \item Reasoning - data collected from sensors then used by the reasoning layer to make decisions about what action the system should perform.
    \item Acting - is used to control the environment by performing specific actions like turning on/off the light or increasing the temperature in the room. 
\end{itemize}
\par In recent years it has become a common practice to combine home automation systems with digital personal assistants. Many big companies have presented their solutions to the market like Google with Google Home, Apple with HomeKit and Amazon with Alexa. All these projects use natural language queries from a user to control the home automation. Amazon Alexa is a good example of a Personal Assistant that can have its tasks extended using different modules such as home automation, news or video handling and more. It provides Alexa Smart Home Skill API for developers to setup automation system and extend the functionality \cite{alexa}. All that developer needs to do is to add new functionality  to specify what type of actions are possible and define the actions which can be invoked using natural language.

\section{Requirements}
The following subsection covers the list of requirements that our prototype of DPA with Smart Home automation extension should satisfy. These requirements were chosen by comparing the already existing Digital Personal Assistance solutions to those of our DPA to identify the minimal, most common set of actions allowing the user to start using the system. The requirements presented in tables FR1 to FR7 are supported by the system while the creation of custom action scenarios is still under development and left as future work.
\begin{table}[h!]

\begin{tabularx}{\linewidth}{|l| X |}
\hline
\textbf{Requirement ID} & FR-01   \\
\hline
\textbf{Title} & Recognition of actions and questions   \\
\hline
\textbf{Description} &
The system should understand a user's query and distinguish between command and question\\
\hline
\textbf{Priority} & Mandatory                 \\ 
\hline
\textbf{Risk} & Critical                   \\ 
\hline
\end{tabularx}
\end{table}

\begin{table}[h!]

\begin{tabularx}{\linewidth}{|l| X |}
\hline
\textbf{Requirement ID} & FR-02   \\
\hline
\textbf{Title} & Information extraction   \\
\hline
\textbf{Description} &
The DPA should extract numbers, dates and string corresponding to pattern.\\
\hline
\textbf{Priority} & Mandatory                 \\ 
\hline
\textbf{Risk} & Critical                   \\ 
\hline
\end{tabularx}
\end{table}

\begin{table}[h!]

\begin{tabularx}{\linewidth}{|l| X |}
\hline
\textbf{Requirement ID} & FR-03   \\
\hline
\textbf{Title} & Dialog support  \\
\hline
\textbf{Description} &
The DPA should ask a user additional question if the information provided by the user is not enough for the execution of a query. \\
\hline
\textbf{Priority} & Mandatory                 \\ 
\hline
\textbf{Risk} & Critical                   \\ 
\hline
\end{tabularx}
\end{table}

\begin{table}[h!]

\begin{tabularx}{\linewidth}{|l| X |}
\hline
\textbf{Requirement ID} & FR-04   \\
\hline
\textbf{Title} & Sensor information   \\
\hline
\textbf{Description} &
Information, captured by any sensor must be shown in the app. \\
\hline
\textbf{Priority} & Mandatory                 \\ 
\hline
\textbf{Risk} & Critical                   \\ 
\hline
\end{tabularx}
\end{table}

\begin{table}[h!]

\begin{tabularx}{\linewidth}{|l| X |}
\hline
\textbf{Requirement ID} & FR-05   \\
\hline
\textbf{Title} & Actuator use   \\
\hline
\textbf{Description} &
Actuators must be controlled via application.\\
\hline
\textbf{Priority} & Mandatory                 \\ 
\hline
\textbf{Risk} & Critical                   \\ 
\hline
\end{tabularx}
\end{table}

\begin{table}[h!]

\begin{tabularx}{\linewidth}{|l| X |}
\hline
\textbf{Requirement ID} & FR-06   \\
\hline
\textbf{Title} & Anytime access.    \\
\hline
\textbf{Description} &
The system must be at user's disposal at any time. \\
\hline
\textbf{Priority} & Mandatory                 \\ 
\hline
\textbf{Risk} & Critical                   \\ 
\hline
\end{tabularx}
\end{table}

\begin{table}[h!]

\begin{tabularx}{\linewidth}{|l| X |}
\hline
\textbf{Requirement ID} & FR-07   \\
\hline
\textbf{Title} & Excitability  \\
\hline
\textbf{Description} &
The DPA must allow third-party developers to extend its functionality via their devices and applications\\
\hline
\textbf{Priority} & Mandatory                 \\ 
\hline
\textbf{Risk} & Critical                   \\ 
\hline
\end{tabularx}
\end{table}

\newpage
\section{Use cases}
Use cases define the interaction with the DPA for achieving specific goals such as turn on/off activators, display measure of sensors and change of temperature.

\begin{table}[h!]
\begin{tabularx}{\linewidth}{|l| X |}
\hline
\textbf{User Case ID} & UC-01   \\
\hline
\textbf{Title} & Extension  \\
\hline
\textbf{Description} &
I as a developer can integrate with the DPA without changes in DPA’s code so that I can provide extensions of the functionality of the DPA. \\
\hline
\end{tabularx}
\end{table}

\begin{table}[h!]
\begin{tabularx}{\linewidth}{|l| X |}
\hline
\textbf{User Case ID} & UC-02   \\
\hline
\textbf{Title} & Dashboard information  \\
\hline
\textbf{Description} &
I as a user can request information about particular sensor so that I can know the current status of the environment.\\
\hline
\end{tabularx}
\end{table}

\begin{table}[h!]
\begin{tabularx}{\linewidth}{|l| X |}
\hline
\textbf{User Case ID} & UC-03   \\
\hline
\textbf{Title} & Activator control  \\
\hline
\textbf{Description} &
I as a user can turn on and off a presence in the system activators so that I can manage the work of activators.\\
\hline
\end{tabularx}
\end{table}

\section{Architecture}
\subsection{General architecture}
Our prototype ecosystem consists of the DPA and 3-D party applications. The goal of the DPA is to receive, recognize and extract information from user’s request in natural language and then send it in machine understandable format to particular 3-D party app. In its turn, this app has to execute the user’s intent and provide the result of its job.
\par A 3-D party programmer can easily integrate his/her application with the DPA by adding description (Listing \ref{listing:app-json}) of intents which the app can handle and what intents' parameters are expected. Important to note is that the 3rd party system can be fully autonomous and work independently from the DPA.

\begin{lstlisting}[frame=single,language=json,caption=Description of an application, label=listing:app-json]
{ "name": "Home",
  "description": "Home management app",
  "type": "RemoteApp",
  "url": "http://127.0.0.1:7878",
  "intents": [{
    "name": "Set temperature",
    "samples":[
        "Change temperature",
        "Set up 5 degree"]
    "key_phrases": ["set", "set up"],
    "parameters": [{
          "name": "temperature",
          "data_type": "Number",
          "obligatory": true,
          "question": "What temperature
                       I should setup?"}]}
\end{lstlisting}

\subsection{DPA architecture}
We are developing the DPA with the idea that any programmer can add easily support of his/her native language. For this purpose, we developed API for a language model. Each language is a module which has to handle particular language and provide information about the user’s request in uniform format for all languages. The format includes information about separating text on tokens and their description such as part of speech, named type and so on.
\par In the first stage, we implemented the support of English. The language model for it is implemented with using CoreNLP project. CoreNLP separates text on tokens and does PoS and NER tagging.
\par The platform can extract three data type from a text. They are date, number and string which correspond to a pattern. At the current stage, the DPA expects that a request does not contain more than one value for each data type. This restriction allows to solve a slot filling problem in a straightforward way.
\par After parsing the user's utterance, the DPA begins to look for suitable intent. First of all, the assistant calculates the WMD for the incoming request and all samples of all intents. If a minimal distance for a sample is lower then the predefined threshold then it stops the search. If the search by samples can not find a close sample then the assistant tries to find intent by key phrases. If even after it the DPA does not find anything, we notify the user that we do not recognize his/her request.
\par In case of success search, the DPA takes meta information about expected arguments for the intent and based on it, tries to fill slots of intent by values found in the user request. If the request does not contain a value for an obligatory slot, the DPA asks predefined additional question for the slot. When all obligatory slots will be filled they will be passed to intent handler. After obtaining the handler execution result, the assistant displays it to user.
\par In order to interact with the DPA, the user just needs to type a message to it thought Telegram messanger.

\subsection{Smart home architecture}
Building smart home environment requires a number of sensors and actuators as well \cite{gusmanov2016jolie} as a Digital Personal Assistant which will allow us to control all the hardware components and collect the data. Arduino is an open-source electronic platform that is based on an easy-to-use hardware components \cite{Arduino_Introduction}. It provides huge variety of inexpensive hardware devices such as sensors, actuator, controllers and many other things that can be used to build home automation. Another big advantage is that all hardware components are compatible with each other.  \\

In our implementation we used:
\begin{itemize}
    \item Esp8266 (ESP) esp12-e programmable and Arduino compatible WiFi board.
    \item DHT11 humidity and temperature sensor
    \item Light intensity sensor
    \item Set of relays to turn on/off appliances such as light, heater. 
\end{itemize}

\begin{figure}[h]
\centerline{\includegraphics[scale =0.5]{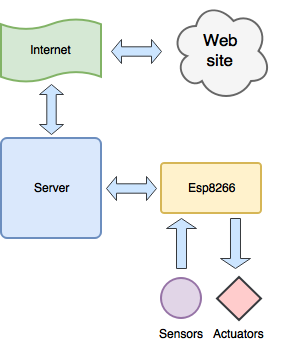}}
\caption{[General Architecture of Smart Home Assistant]}
\label{fig:General_Architecture_of_Smart_Home}
\end{figure}
The automation system is fully self-contained. Figure \ref{fig:General_Architecture_of_Smart_Home} shows the architecture of Smart Home Assistant, the system consists of two main parts: 
\begin{itemize}
    \item The first part is the python server that controls all the logic and stores the data, it also acts like a DPA that can be accessed through a web browser. Using the assistant, the user can issue commands or monitor the current state of the environment.
    \item The second part of the system is Arduino based microprocessor with WiFi module on the board that was programmed to retrieve data from sensors and send them to the server with fixed time intervals; it also receives commands from the server to turn on/off relays of specific devices.
\end{itemize}

\begin{figure}[h!]
\centerline{\includegraphics[scale=0.4]{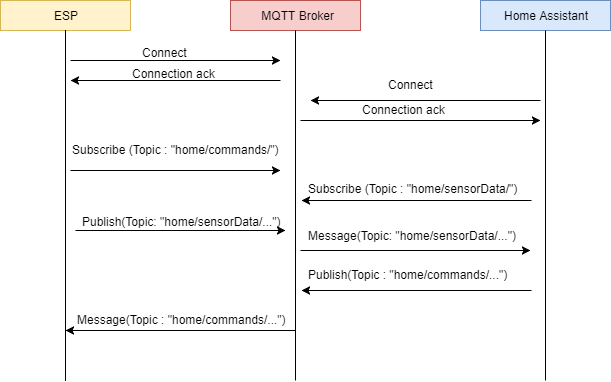}}
\caption{[MQTT messaging sequence diagram]}
\label{fig:MQTT_messaging}
\end{figure}

The Communication of the server and Arduino is done by MQTT protocol that is designed especially for IoT devices with low energy consumption, limited computational power and low-bandwidth  connections. It is very lightweight in comparison to regular HTTP requests and contains only useful information. MQTT requires 3 main entities: Publisher (one that sends the information), Subscriber (one that subscribed to the publisher to receive information) and Broker (controls all the messaging process, making sure that every subscriber receives a message send by the publisher). The same entity can be a publisher and a subscriber at the same time, so in our system, we have a Server that is subscribed to ESP to collect data from sensors and ESP that is subscribed to the server receiving commands. We used Mosquitto open source broker to manage message passing. Figure \ref{fig:MQTT_messaging} represents how message passing between ESP and Smart Home Assistant is done it the project.
 \\
\par There are two main functions in the communication between the Home Assistant and IoT devices. The first is collecting information from sensors and the second is to send commands to actuators. Initially, both ESP and the Home Assistant are connected to the broker. After acknowledgements of the connections, ESP which controls all the IoT devices, subscribes to a specified topic named "home/commands" which is used by the Assistant to issue the commands. The Home Assistant on the other hand subscribes to a topic "home/sensorData" to which ESP will send information from the sensors. Now, when ESP is ready to send data, it publishes the message to the broker, using a pre-defined topic ("home/sensorData/.values."). After publishing the message, Broker takes all the responsibility to transfer the message to all Subscribers of a given topic, it attempts to deliver the message until it is successfully delivered to all subscribers. The same procedure happens when the Home Assistant is ready to send the command to ESP.

\section{Evaluation}
Set of test cases were developed in order to evaluate the system. These test cases are intended to check the reliability of the system as well as its functionality. All the test cases are presented in tables TC-1 to TC-4 were successfully passed. The automation system was tested for a week in real world scenario in a room at home. During the test, the time system always performed the user's commands correctly and had  $100\%$ up-time.
\par The resulting system allows the user to make requests in natural language text. This text is managed by the DPA to retrieve the intent of the user, meaning that it finds what the 3-D party application should do. In our case, the 3-D party application is Smart Home, so if a user orders, "Home, turn off the light", the DPA recognizes that the command should be executed by the Smart Home, the action is "turn off" and the object to which we are to apply this order is “light”. The DPA then creates a JSON out of these variables and sends them to the server of the smart home via POST request. The Home Assistant now just executes the received commands and turns off the lights.

\par The Smart Home can be controlled either via natural language commands or through the web interface. The system satisfies most of the functional requirements that were described in the Requirements section.

\begin{table}[h!]
\centering
\begin{tabularx}{\linewidth}{|c| X |}
\hline
\textbf{Test Case ID}     & TC01                                         \\ \hline
\textbf{Description}      & To verify that users queries are recognized correctly  \\ \hline
\textbf{Reqs}             & FR01,FR02,FR03                                                        \\ \hline
\textbf{Test Procedure}   &       
1.Input query "Home, turn on the lights" in DPA \\ \hline
\textbf{Expected Result} &       
1.Resulting JSON file should specify that the Home Assistant was chosen as a delegate, Light is the entity that should be affected, and turn on is the commands.
\\ \hline
\textbf{Status}           &                                     Success                        \\ \hline
\end{tabularx}
\end{table}

\begin{table}[h!]
\begin{tabularx}{\linewidth}{|c| X |}
\hline
\textbf{Test Case ID}     & TC02                                         \\ \hline
\textbf{Description}      & \par To verify that ESP successfully controls relays \\ \hline
\textbf{Reqs}             & FR05                                                        \\ \hline
\textbf{Test Procedure}   &       1.Connect ESP with sensors to PC 
\par 2.Open Serial Monitor to send commands to ESP
\par 3.Send command to change I/O port connected to relay to HIGH.
\par 4.Send command to change I/O port connected to relay to LOW.\\ \hline
\textbf{Expected Result} &       
1.Setting I/O port to HIGH should switch relay and turn on status led.
\par 2.Setting I/O port to LOW should switch relay and turn off status led.\\ \hline
\textbf{Status}           &                                     Success                        \\ \hline
\end{tabularx}
\end{table}

\begin{table}[h!]
\centering
\begin{tabularx}{\linewidth}{|c| X |}
\hline
\textbf{Test Case ID}     & TC03                                         \\ \hline
\textbf{Description}      & To verify that MQTT broker works properly and ESP can successfully send sensor information through MQTT \\ \hline
\textbf{Reqs}             & FR04,FR05,FR06                                                        \\ \hline
\textbf{Test Procedure}   &       
1.Upload test program to ESP that will connect to MQTT broker and send data from sensors every 2 seconds using "home/sensorData" topic.
\par 2.Open terminal on PC running MQTT and connect to MQTT broker.
\par 3.Subscribe to topic "home/sensorData"\\ \hline
\textbf{Expected Result} &       
1.Every 2 seconds the terminal should show new data from the sensors.
\\ \hline
\textbf{Status}           &                                     Success                        \\ \hline
\end{tabularx}
\end{table}

\begin{table}[h!]
\centering
\begin{tabularx}{\linewidth}{|c| X |}
\hline
\textbf{Test Case ID}     & TC04                                         \\ \hline
\textbf{Description}      & To verify that the DPA successfully delegates commands execution to the Home Assistant and these commands are successfully performed \\ \hline
\textbf{Reqs}             & FR01,FR02,FR03,FR05                                                       \\ \hline
\textbf{Test Procedure}   &       
1. Connect the DPA with the Home Assistant
\par 2. Input query "Home, turn on the lights" in the DPA
\par 3. Input query "Home, turn off the light" \\ \hline
\textbf{Expected Result} &       
1. Relay specified for the light should be switched on after the first query.
\par 2. Relay specified for the light should be switched off after the second query
\\ \hline
\textbf{Status}           &                                     Success                        \\ \hline
\end{tabularx}
\end{table}

Examples in tables EX1 to EX4 show what kind of queries the system is able to recognize and show the expected results of entity recognition for each case.

\begin{table}[h!]
\centering
\begin{tabularx}{\linewidth}{|c| X |}
\hline
\textbf{Example ID}     & EX01                                         \\ \hline
\textbf{Dialog}      &
\par User: Home, turn off
\par DPA: What should I turn off?
\par User: the computer  \\ \hline
\textbf{Expected JSON}    &\{"AppName": "Home", "Intent":"Turn off", "object": "the computer"\} \\ \hline
\textbf{Status}           &                                     Success                        \\ \hline
\end{tabularx}
\end{table}

\begin{table}[h!]
\centering
\begin{tabularx}{\linewidth}{|c| X |}
\hline
\textbf{Example ID} & EX02 \\ \hline
\textbf{Dialog}     & User: Calendar, remind 16th of November to meet Sasha \\ \hline
\textbf{Expected JSON} & \{"AppName": "Calendar", "Intent":"Create remind", "Subject": "meet Sasha", "Date": "16th of November"\} \\ \hline
\textbf{Status}           & Success   \\ \hline
\end{tabularx}
\end{table}

\begin{table}[h!]
\centering
\begin{tabularx}{\linewidth}{|c| X |}
\hline
\textbf{Example ID} & EX03 \\ \hline
\textbf{Dialog}     & 
\par User: Calendar, remind
\par DPA: When should I remind you?
\par User: on Monday
\par DPA: What should I remind you?
\par User: meet Sasha on the airport
\\ \hline
\textbf{Expected JSON} & \{"AppName": "Calendar", "Intent":"Create remind", "Subject": "meet Sasha on the airport", "Date": "Monday"\} \\ \hline
\textbf{Status}           & Success   \\ \hline
\end{tabularx}
\end{table}

\begin{table}[h!]
\centering
\begin{tabularx}{\linewidth}{|c| X |}
\hline
\textbf{Example ID}     & EX04                                         \\ \hline
\textbf{Dialog}      &
\par User: Home, turn off air conditioning \\ \hline
\textbf{Expected JSON}    &\{"AppName": "Home", "Intent":"Turn off", "object": "air conditioning"\}  \\ \hline
\textbf{Status}           &                                     Success                        \\ \hline
\end{tabularx}
\end{table}

\section{Conclusion}
In this paper, we have introduced our prototype of an open source DPA. The system consists of two main components:
\begin{itemize}
    \item Conversational Interface that is able to translate natural human language to a set of machine commands.
    \item Smart Home agent that executes the commands issued by the user to control home appliances as well as to provide the possibility to monitor data from the sensors deployed in a house.
\end{itemize}

We have demonstrated the functionality of the DPA and the Smart Home by showing their interaction with a real person. As future work, we will extend the Conversational interface to include a dialog support system, introduce new language modules and improve the slot filling functionality. Regarding the Smart Home agent, we intend to work on the possibility to create custom scenarios, improve data analysis, and concentrate on the safety aspects, to make the system more intelligent and safe. Future work will have to take into account the functioning of Personal Assistants in different scenarios, for example automotive \cite{gmehlich2013towards}, and how to re-engineer it in order to be deployed in a flexible and continuous fashion, for example following the microservice paradigm \cite{Dragoni2017, DragoniLLMMS17}.

\bibliographystyle{IEEEtran}
\bibliography{IEEEabrv,conference_071817} 

\end{document}